\documentclass[12pt]{article}

\usepackage{amsmath}
\usepackage{amssymb}
\usepackage[dvips]{graphicx}
\usepackage[dvips]{psfrag}
\usepackage{cite}

\makeatletter
 
  \@addtoreset{equation}{section}
 \makeatother

\newcommand {\n}{\nonumber \\}
\newcommand {\tr}{\mbox{tr}}

\begin{document}
\setlength{\oddsidemargin}{0cm}
\setlength{\baselineskip}{7mm}

\begin{titlepage}
\begin{normalsize}
\begin{flushright}
\begin{tabular}{l}
KUNS-2090\\
August 2007
\end{tabular}
\end{flushright}
  \end{normalsize}

~~\\

\vspace*{0cm}
    \begin{Large}
       \begin{center}
         {Perturbative Vacua from IIB Matrix Model}
       \end{center}
    \end{Large}
\vspace{1cm}

\begin{center}
           Hikaru K{\sc awai}$^{a,b}$\footnote
            {
e-mail address : hkawai@gauge.scphys.kyoto-u.ac.jp}
           {\sc and}
           Matsuo S{\sc ato}$^{a}$\footnote
           {
e-mail address : satomat@gauge.scphys.kyoto-u.ac.jp}\\
      \vspace{1cm}
       
        $^{a}$ {\it Department of Physics, Kyoto University, Kyoto 606-8502, Japan}\\

        $^{b}$ {\it Theoretical Physics Laboratory}\\
                {\it The Institute of Physical and Chemical Research (RIKEN)}\\
                {\it Wako, Saitama 351-0198, Japan}
\end{center}

\hspace{5cm}

\begin{abstract}
\noindent
It has not been clarified whether a matrix model can describe various vacua of string theory. In this paper, we show that the IIB matrix model includes type IIA string theory. In the naive large N limit of the IIB matrix model, configurations consisting of simultaneously diagonalizable matrices form a moduli space, although the unique vacuum would be determined by complicated dynamics. This moduli space should correspond to a part of perturbatively stable vacua of string theory. Actually, one point on the moduli space represents type IIA string theory. Instead of integrating over the moduli space in the path-integral, we can consider each of the simultaneously diagonalizable configurations as a background and set the fluctuations of the diagonal elements to zero. Such procedure is known as quenching in the context of the large N reduced models. By quenching the diagonal elements of the matrices to an appropriate configuration, we show that the quenched IIB matrix model is equivalent to the two-dimensional large N $\mathcal{N}=8$ super Yang-Mills theory on a cylinder. This theory is nothing but matrix string theory and is known to be equivalent to type IIA string theory. As a result, we find the manner to take the large N limit in the IIB matrix model. 
\end{abstract}

\vfill
\end{titlepage}
\vfil\eject

\setcounter{footnote}{0}

\section{Introduction}
\setcounter{equation}{0}

The IIB matrix model is one of the proposals for non-perturbative string theory \cite{IKKT}. In the original interpretation, it naturally describes type IIB string theory. In \cite{FKKT, AIKKTT}, the light-cone Hamiltonian for type IIB string field theory is derived from Schwinger-Dyson equations for Wilson loops. On the other hand, matrix string theory describes type IIA string theory \cite{Motl, BS, DVV, DM}. In this theory, the diagonal elements of the eight scalars form coordinates of the light-cone strings. This interpretation correctly reproduces the world-sheet action and the joining and splitting of type IIA strings \cite{Wynter, 2BBN}. However, it has not been clarified whether a matrix model can produce two or more perturbative string theories, although a non-perturbative string theory should include all perturbative vacua.

In the large N limit of the IIB matrix model, configurations of simultaneously diagonalizable matrices form a moduli space at least in the one-loop level. Here we discuss stability of such configurations. In a naive reduced model given by $I=-\frac{1}{4g^2}\tr[A_{\mu}, A_{\nu}]^2$ ($\mu=1, \cdots, D$), the one-loop effective action for the diagonal elements $p_{\mu}^{i}$ is given by $\displaystyle S=(D-2) \sum_{i<j} \log \left((p_{\mu}^{i}-p_{\mu}^{j})^2 \right)$. Therefore, if $D>2$ extended configurations of the diagonal elements are unstable because they collapse to a point. In a supersymmetric case, if one ignores the diagonal elements of fermionic matrices, the one-loop effective action for the diagonal elements of bosonic matrices is given by $\displaystyle S=(D-2-d_F) \sum_{i<j} \log \left((p_{\mu}^{i}-p_{\mu}^{j})^2 \right)=0$, and there is no force between them. However, one cannot ignore the diagonal elements of fermions when the dimensions of a theory is less than one. In fact, the one-loop effective action for the diagonal elements of both the bosonic and fermionic matrices in the IIB matrix model is given by 
$ \displaystyle
S(p, \xi)=\sum_{i<j} \tr \left(\frac{S_{(i,j)}^4}{4} +\frac{S_{(i,j)}^8}{8}\right),
$
where 
$
\left(S_{(i,j)}\right)_{\mu,\nu}
=
(\bar{\xi}^{(i)} - \bar{\xi}^{(j)})\Gamma^{\mu\rho\nu}( {\xi}^{(i)} -{\xi}^{(j)} )\frac{p_{\rho}^{(i)}-p_{\rho}^{(j)}}{\left((p_{\lambda}^{(i)} -p_{\lambda}^{(j)})^2\right)^2}
$ \cite{AIKKT}.
By integrating out $\xi$, we have a complicated interaction among $p_{\mu}^i$
$\displaystyle \exp\left(-S(p)\right)=\int \prod_i d^{16} \xi^{(i)} \exp\left(-S(p,\xi)\right)=\int \prod_{i=1}^{N} d^{16} \xi^{(i)}\prod_{i<j}\left(1+a\tr(S_{(i,j)}^4)+b\tr(S_{(i,j)}^8)\right),
$
which is estimated as follows. In the last expression, for each pair of i and j we have three choices $1$, $a\tr(S_{(i,j)}^4)$, $b\tr(S_{(i,j)}^8)$, which carry 0, 8, 16 powers of $\xi$, respectively. Because we have 16N dimensional fermionic integral $\displaystyle \int \prod_{i=1}^{N} d^{16} \xi^{(i)}$, the number of factors other than 1 should be less than or equal to 2N. Therefore, the effective action for $p_{\mu}^i$ is expressed as a sum of terms consisting of less than or equal to $2N$ factors:
$$
\exp\left(-S(p)\right)
=
\sum_{\mbox{various terms}} f(\frac{p_{\rho}^{(i)}-p_{\rho}^{(j)}}{\left((p_{\mu}^{(i)} -p_{\mu}^{(j)})^2\right)^2})f'(\frac{p_{\rho}^{(i')}-p_{\rho}^{(j')}}{\left((p_{\mu}^{(i')} -p_{\mu}^{(j')})^2\right)^2}) \cdots \sim\exp(O(N)), 
$$
where $f$, $f'$, $\cdots$ are polynomials. This should be compared to the effective action in the bosonic case, which is of order 
$
\exp(O(N^2)).
$
We see that supersymmetry reduces the attractive force by order $1/N$ at least in the one-loop level. If this is true to all orders, in the large N limit any of the simultaneously diagonalizable configurations is stable and represents an independent vacuum as in the case of a moduli of scalar fields in the ordinary field theory 
\footnote
{However, such configurations would become unstable if we take the $1/N$ corrections into account. This would correspond to instability of perturbative vacua of string theory when non-perturbative corrections are included.
}
. Contributions from such vacua are approximately of the same order in the path-integral of the IIB matrix model.

Instead of integrating over the moduli space, we can consider each of the simultaneously diagonalizable configurations as a background and set the fluctuations of the diagonal elements to zero. Such procedure is known as quenching in the context of the large N reduced model. In this paper, we show that type IIA string emerges as a vacuum of the IIB matrix model, if we introduce such interpretation. More precisely, we show that the IIB matrix model quenched appropriately is equivalent to matrix string theory, which gives type IIA string theory. The moduli space of the IIB matrix model also includes other perturbative vacua. For example, we can show that the IIB matrix model quenched in another way gives four-dimensional $\mathcal{N}=4$ super Yang-Mills theory, thus produces type IIB string theory on the $AdS_5 \times S^5$.

The organization of this paper is as follows. In section 2, we show that by introducing a proper quenching the IIB matrix model becomes equivalent to the two-dimensional $\mathcal{N}=8$ super Yang-Mills theory on a cylinder, which is nothing but matrix string theory. In section 3, we find relations between the coupling constant $g_{IIB}$ of the IIB matrix model and the string coupling $g_s$, through matrix string theory. As a result, we find how we should take the large N limit in the IIB matrix model. In section 4, we summarize and discuss our results. 

\vspace{1cm}

\section{Gauge Theories from IIB Matrix Model}
\setcounter{equation}{0}

In general, zero-dimensional matrix models are obtained by dimensional reduction of gauge theories. Such models can reproduce the gauge theories in the large N limit by quenching the diagonal elements to uniformly distributed values \cite{EK, Parisi, BHN, GK, KN}. If we consider Feynman diagrams, the diagonal elements behave as the momenta in the gauge theories, and sums over indices become integrals over the momenta in the large N limit. In this way, the quenched matrix models and the gauge theories are equivalent. Therefore, the IIB matrix model can produce the maximally supersymmetric large N gauge theories by quenching the diagonal elements of some matrices to uniformly distributed values and those of the other matrices to zero. As a generalization, if we quench the diagonal elements to discrete values instead of continuous ones, the matrix model corresponds to a toroidally compactified gauge theory, because the discrete momenta are conjugate to compactified coordinates.

In the following, we show that the IIB matrix model quenched in an appropriate way gives the two-dimensional $\mathcal{N}=8$ super Yang-Mills theory on a cylinder, whose bosonic sector is given by 
\begin{equation}
S_{2D}=\frac{N}{\lambda}\int^{\infty}_{-\infty} dt \int^L_0 dx \, \tr \left(\frac{1}{4} (\mathcal{F}_{\mu\nu})^2+\frac{1}{2}(\mathcal{D}_\mu \mathcal{A}^I)^2 - \frac{1}{4} [\mathcal{A}^I, \mathcal{A}^J]^2 \right),
\label{2dimSYM}
\end{equation}
where $\mu, \nu =1, 2$ and $I=3, \cdots, 10$.

\begin{figure}[htbp]
\begin{center}
\psfrag{p1}{$p^1$}
\psfrag{p2}{$p^2$}
\psfrag{0}{$0$}
\includegraphics[height=5cm, keepaspectratio, clip]{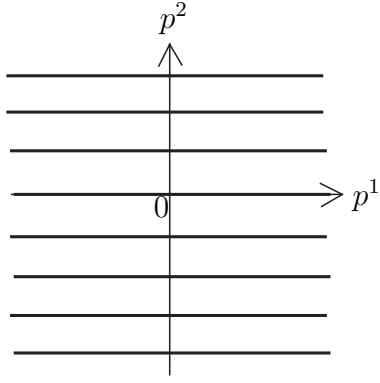}\end{center}
\caption{Uniform distribution of the diagonal elements}
\label{distribution}
\end{figure}

We start with the IIB matrix model, whose bosonic part is given by
\begin{equation}
S=-\frac{1}{4 g_{IIB}^2}\tr([A^M, A^N]^2). \label{IIBaction}
\end{equation}
Let $p^M_i$ $(i=1, \cdots, N)$ be the diagonal elements of the matrices $A^M$. We regard $(p_i^1, \cdots, p_i^{10})$ as a ten-dimensional vector for each $i$. We assume such vectors distributed uniformly in 
\begin{equation}
\{
(p, \frac{2\pi}{L}n, 0, \cdots, 0)| 
p \in \mathcal{R}, \,\,
n \in \mathcal{Z}, \,\,
-\frac{\Lambda}{2} < p < \frac{\Lambda}{2}, \,\,
-\frac{\Lambda}{2} < \frac{2\pi}{L}n < \frac{\Lambda}{2}
\}, \label{region}
\end{equation}
as Fig. \ref{distribution} and set the fluctuations of the diagonal elements to zero.

By introducing the 't Hooft coupling
\begin{equation}
\lambda = Ng_{IIB}^2 (\frac{2\pi}{\Lambda})^2, \label{relation}
\end{equation}
(\ref{IIBaction}) is rewritten as
\begin{equation}
S=-\frac{N}{4 \lambda} \left(\frac{2\pi}{\Lambda}\right)^2 \tr([A^M, A^N]^2).
\label{quenched}
\end{equation}
We then expand the matrices as 
\begin{eqnarray}
A^{\mu}&=&p^{\mu}+ a^{\mu} \quad (\mu=1, 2) \n
A^{I}&=& a^{I} \quad (I=3, \cdots, 10), \label{fields}
\end{eqnarray}
where all the diagonal elements of $a^{\mu}$ and $a^{I}$ are fixed to zero. In order to obtain the Feynman rule for this action, we choose a gauge fixing condition as $F(a_{\mu})=[p^{\mu}, a_{\mu}]=0$. Then, we have a gauge fixing term $\frac{N}{\lambda}\left(\frac{2\pi}{\Lambda}\right)^2 \frac{1}{2} \tr([p^{\mu}, a_{\mu}]^2)$ and a ghost term $\frac{N}{\lambda}\left(\frac{2\pi}{\Lambda}\right)^2 \tr (\bar{c} [p^{\mu}, [p_{\mu} +  a_{\mu}, c]])$ in the Feynman gauge. The total action is given by 
\begin{eqnarray}
S &=& \frac{N}{\lambda}\left(\frac{2\pi}{\Lambda}\right)^2 
\Biggl(
 \frac{1}{2}(p^i_{\mu}-p^j_{\mu})^2 a_{M}^{ij} a^{M ji}
+(p^i_{\mu}-p^j_{\mu})^2 \bar{c}^{ij} c^{ji} \n
&& \qquad \qquad - \left( (p^k_{\mu}-p^i_{\mu})- (p^j_{\mu}-p^k_{\mu}) \right)
a^{\mu ij} a_{M}^{jk} a^{M ki}  \n
&& \qquad \qquad -\frac{1}{2}
(a_M^{ij} a_N^{jk} a^{M kl} a^{N li}- a_M^{ij} a^{M jk} a_N^{kl} a^{N li}) \n
&& \qquad \qquad + (p^i_{\mu}-p^j_{\mu}) (\bar{c}^{ij} c^{jk} a^{\mu ki}
- \bar{c}^{ij} a^{\mu jk} c^{ki} )
\Biggr), \label{quenched}
\end{eqnarray}
where $i, j, k, l = 1, \cdots, N$, $\mu=1,2$, $I=3, \cdots, 10$ and $M=1, \cdots, 10$. 
From this form, we can see that there is a one-to-one correspondence between 
the Feynman rule for this action and that for (\ref{2dimSYM}). For example, the propagators are given by
\begin{eqnarray}
\langle a^{M}_{ij}a^{N}_{kl}\rangle  &=&\eta^{MN}\delta_{il}\delta_{jk} \frac{\lambda}{N}\left(\frac{\Lambda}{2\pi}\right)^2 \frac{1}{(p_{\mu}^i-p_{\mu}^j)^2} \n
\langle c_{ij} \bar{c}_{kl}\rangle  &=&\delta_{il}\delta_{jk} 
\frac{\lambda}{N}\left(\frac{\Lambda}{2\pi}\right)^2 \frac{1}{(p_{\mu}^i-p_{\mu}^j)^2},
\label{SYMpropagator}
\end{eqnarray}
where
$(p^i_{\mu}-p^j_{\mu})^2=(p^i-p^j)^2+\left(\frac{2\pi}{L}\right)^2(n^i-n^j)^2$.
Note that the diagonal elements $\langle a^{\mu}_{ii} a^{\nu}_{jj}\rangle$, $\langle a^{I}_{ii} a^{J}_{jj} \rangle$ and $\langle c_{ii} \bar{c}_{jj} \rangle$ do not appear because of the quenching.

In addition to the one-to-one correspondence between the Feynman rules, we can show that in the large N limit the free energy of the matrix model equals that of the Yang-Mills theory per volume $(\frac{2 \pi}{\Lambda})^2$. If we assume that $\Lambda$ is finite, planer diagrams dominate when $\lambda$ is kept fixed both in the matrix model and Yang-Mills theory. Therefore, the two theories are equivalent under the condition
\begin{equation}
\lambda \mbox{ and } \Lambda \mbox{: fixed.} \label{planer}
\end{equation}

\begin{figure}[htbp]
\begin{center}
\includegraphics[height=5cm, keepaspectratio, clip]{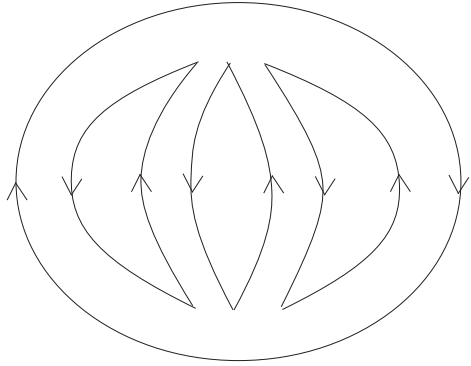}\end{center}
\caption{A planar contribution to the free energy}
\label{bubble}
\end{figure}

As an example, we compare the values of the Feynman diagram depicted in Fig. \ref{bubble}. 
On the matrix model side, we have
\begin{equation}
F_{M}=C
\left(\frac{\lambda}{N}\left(\frac{\Lambda}{2\pi}\right)^2\right)^4
\left(\frac{N}{\lambda}\left(\frac{2\pi}{\Lambda}\right)^2 \right)^2
\sum_{i,j,k,l=1}^N
\frac{1}{(p^i_{\mu}-p^j_{\mu})^2}
\frac{1}{(p^j_{\mu}-p^k_{\mu})^2}
\frac{1}{(p^k_{\mu}-p^l_{\mu})^2}
\frac{1}{(p^l_{\mu}-p^i_{\mu})^2},
\end{equation}
where C is a combinatorial factor. In the large N limit, $\frac{1}{N}\sum_{i=1}^N f(p^i, n^i)$ is replaced by $\frac{1}{\Lambda} \int dq \frac{1}{\Lambda}\frac{2\pi}{L}\sum_{m=-\infty}^{\infty}f(q, m)$ because $p^i$ and $n^i$ are uniformly distributed in $-\frac{\Lambda}{2}<p^i<\frac{\Lambda}{2}$ and $-\frac{\Lambda}{2}\frac{L}{2\pi}<n^i<\frac{\Lambda}{2}\frac{L}{2\pi}$, respectively. Then, $F_{M}$ is rewritten as
\begin{eqnarray}
F_{M}&=&C
\left(\frac{\lambda}{N}\left(\frac{\Lambda}{2\pi}\right)^2\right)^4\left(\frac{N}{\lambda}\left(\frac{2\pi}{\Lambda}\right)^2 \right)^2 \n
&& \times \prod_{a=1}^4 \left(N \frac{1}{\Lambda} \int dq^a \frac{1}{\Lambda}\frac{2\pi}{L}\sum_{m^a=-\infty}^{\infty} \right)
\frac{1}{(q^1_{\mu}-q^2_{\mu})^2}
\frac{1}{(q^2_{\mu}-q^3_{\mu})^2}
\frac{1}{(q^3_{\mu}-q^4_{\mu})^2}
\frac{1}{(q^4_{\mu}-q^1_{\mu})^2},\n
\end{eqnarray}
where $(q^a_{\mu}-q^b_{\mu})^2=(q^a-q^b)^2+\left(\frac{2\pi}{L}\right)^2(m^a-m^b)^2$. If we define $p_{\mu} \equiv q^1_{\mu}-q^2_{\mu}$, $q_{\mu} \equiv q^2_{\mu}-q^3_{\mu}$ and $r_{\mu} \equiv q^3_{\mu}-q^4_{\mu}$, we have $\,q^4_{\mu}-q^1_{\mu}=-p_{\mu}-q_{\mu}-r_{\mu}$, \, and thus $\int dq^4$ and $\sum_{m^4=-\infty}^{\infty}$ are factored out. Finally, we obtain
\begin{eqnarray}
F_{M}&=&C
(N\lambda)^2\left(\frac{2\pi}{\Lambda}\right)^2\frac{1}{L^3}
\int \frac{dp}{2\pi}\int \frac{dq}{2\pi}\int \frac{dr}{2\pi}
\sum_{l,m,n=-\infty}^{\infty}
\frac{1}{p^2+\left(\frac{2\pi}{L}l\right)^2 } 
\frac{1}{q^2+\left(\frac{2\pi}{L}m\right)^2 }
\frac{1}{r^2+\left(\frac{2\pi}{L}n\right)^2 } \n
&&
\times \frac{1}{(p+q+r)^2+\left(\frac{2\pi}{L}(l+m+n)\right)^2 }
\end{eqnarray}

On the other hand, for Yang-Mills theory (\ref{2dimSYM}), we have the free energy per unit volume given by
\begin{eqnarray}
f_{Y}&=&C
(N\lambda)^2\frac{1}{L^3}
\int \frac{dp}{2\pi}\int \frac{dq}{2\pi}\int \frac{dr}{2\pi}
\sum_{l,m,n=-\infty}^{\infty}
\frac{1}{p^2+\left(\frac{2\pi}{L}l\right)^2 } 
\frac{1}{q^2+\left(\frac{2\pi}{L}m\right)^2 }
\frac{1}{r^2+\left(\frac{2\pi}{L}n\right)^2 } \n
&&\times
\frac{1}{(p+q+r)^2+\left(\frac{2\pi}{L}(l+m+n)\right)^2 }.
\end{eqnarray}
Therefore, we have 
\begin{equation}
F_M=\left(\frac{2\pi}{\Lambda}\right)^2 f_{Y}.
\end{equation}

\section{Type IIA String Theory from IIB Matrix Model}
\setcounter{equation}{0}

We have shown that the IIB matrix model quenched in a proper way is equivalent to the two-dimensional $\mathcal{N}=8$ super Yang-Mills theory on a cylinder, which can be regarded as matrix string theory. Therefore, type IIA string theory emerges as a vacuum of the IIB matrix model. This vacuum is specified by two parameters $\Lambda$ and $L$ as in (\ref{region}). In this section, we discuss how the string coupling $g_s$ is expressed in terms of $g_{IIB}$, $\Lambda$ and $L$ and how we should take a large N limit to obtain type IIA string theory. 

In order to compare (\ref{2dimSYM}) with matrix string theory, we introduce the following redefinition:
\begin{subequations}
\begin{eqnarray}
&&\tau=\frac{2\pi}{L}t , \quad
\sigma=\frac{2\pi}{L}x, \quad \label{a2relation}
\\
&&g_s^2=\frac{(2\pi)^3 N}{\lambda L^2}, \quad \label{b2relation}
\\
&&\tilde{\mathcal{A}_{\mu}}=\frac{L}{2\pi}\mathcal{A}_{\mu}, \quad
X^I=\frac{L}{2\pi}g_sl_s\mathcal{A}^I, \label{c2relation}
\end{eqnarray}
\end{subequations}
where $l_s$ is the string scale. Then, (\ref{2dimSYM}) becomes the well-known form of matrix string theory.
\begin{eqnarray}
S_{MS}&=&\frac{1}{2\pi}\int^{\infty}_{-\infty} \!\! d\tau \int^{2\pi}_0 \!\! d\sigma \, 
\tr 
\Biggl(
\frac{1}{4} g_s^2 \left( \tilde{\mathcal{F}}_{\mu\nu} \right) ^2+\frac{1}{2 l_s^2} \left( \tilde{\mathcal{D}}_\mu X^I \right) ^2 - \frac{1}{4} \frac{1}{g_s^2 l_s^4} [X^I ,X^J]^2 \Biggr). \n \label{matrixstring}
\end{eqnarray}
Note that because of the redefinition of the world-sheet coordinates (\ref{a2relation}), the UV cut-off for this action is given by \begin{equation}
\Lambda_{\mbox{matrix string}} = \frac{L}{2\pi} \Lambda. \label{UVcut}
\end{equation}

From (\ref{a2relation}), (\ref{b2relation}) and (\ref{c2relation}), we can obtain the relation between the IIB matrix model and type IIA string theory. First, substituting (\ref{relation}) to (\ref{b2relation}), $g_s$ is expressed in terms of the parameters of the matrix model,
\begin{equation}
g_s=\frac{\sqrt{2\pi}}{L} \frac{\Lambda}{g_{IIB}}. \label{constants}
\end{equation}
Next, we relate the string coordinates to matrices. The diagonal elements of $X^I$ are string coordinates in the light-cone gauge \cite{DVV}. The relation between $X^I$ and $\mathcal{A}^I$ is given by the second equation of (\ref{c2relation}). When showing the equivalence of (\ref{2dimSYM}) and (\ref{quenched}), we have identified ${\mathcal{A}}^I$ with $a^I$, which is nothing but $A^I$ by (\ref{fields}). Thus we find that the $I$-th components of $A_M^{phys}$ defined by 
\begin{equation}
A_M^{phys}=sA_M, \quad s=\frac{L}{2\pi}g_sl_s \label{scale1}
\end{equation}
represent the string coordinates. In other words, an operator as
\begin{equation}
\tr\left(P(A_M^{phys}) e^{ik^I A_I^{phys}}\right)
\end{equation}
corresponds to the emission vertex of a state with momentum $k^I$, where $P$ is an appropriate polynomial.

Now we discuss the manner to take the large N limit. From (\ref{relation}) and (\ref{constants}) we find that $g_{IIB}$ and $L$ should be tuned in the large N limit as\begin{eqnarray}
g_{IIB}&=&\frac{\Lambda}{2\pi}\sqrt{\frac{\lambda}{N}} \n
L&=&\frac{(\sqrt{2\pi})^3}{g_s}\sqrt{\frac{N}{\lambda}} \label{prepre}
\end{eqnarray}
Note that we have assumed that $\lambda$ and $\Lambda$ is kept finite (\ref{planer}). In this limit, the UV cut-off $\Lambda_{\mbox{matrix string}}$ (\ref{UVcut}) goes to infinity as $O(N^{\frac{1}{2}})$, which guarantees type IIA string theory is produced from the IIB matrix model.

The IIB matrix model has a freedom of redefining the overall scale of the matrices. That is, the form of the action (\ref{IIBaction}) is unchanged under
\begin{equation}
A^M=\kappa A^{'M} \quad g_{IIB}= \kappa^2 g'_{IIB} \label{redef},
\end{equation}
where $\kappa$ is a constant. Here we consider $A_M^{phys}$ defined in (\ref{scale1}) as fundamental variables. Then the parameters $g_{IIB}$, $\Lambda$ and $L$ become $g_{IIB}^{phys}$, $\Lambda^{phys}$ and $L^{phys}$ which are given by 
\begin{eqnarray}
g_{IIB}^{phys}&=&s^2g_{IIB} \n 
\Lambda^{phys}&=&s \Lambda \n
L^{phys}&=&s^{-1} L. \label{pre}
\end{eqnarray}
Here the second and third equations follow from the fact that $\Lambda$ and $L^{-1}$ specify the eigenvalue distributions of $A^{\mu}$, and thus scale in the same way as $A^{\mu}$. By substituting (\ref{prepre}) to (\ref{pre}), we find how the large N limit should be taken in order for $A_I^{phys}$ to represent the string coordinates:
\begin{subequations}
\begin{eqnarray}
&&g_{IIB}^{phys}=C l_s^2 N^{\frac{1}{2}} \label{tune1}
\\
&&\Lambda^{phys}=\sqrt{2\pi} C l_s N^{\frac{1}{2}} \label{tune2}
\\
&&L^{phys}=\frac{2\pi}{g_s l_s}, \label{tune3}
\end{eqnarray}
\end{subequations}
where $C$ is defined by
\begin{equation}
C=\frac{\Lambda}{\sqrt{\lambda}}.
\end{equation}
Using (\ref{relation}), we can rewrite $C$ as $\frac{1}{2\pi \sqrt{N}} \frac{\Lambda^2}{g_{IIB}}$, which indicates that $C$ is invariant under the redefinition (\ref{redef}).

So far we have discussed the leading order in the large N limit. As we discussed in the introduction, although we can freely fix eigenvalues of the matrices by hand in this order, 1/N corrections should determine their distribution dynamically \cite{AIKKT, KNS, AW, NS, KKKMS}. In this case, the square of the range of the eigenvalue distribution $\Lambda$ is expressed by a function $f$ as 
\begin{equation}
\Lambda^2 = g_{IIB} f(N) \label{dym}
\end{equation}
because it is given by
\begin{eqnarray} 
\left\langle
\frac{1}{N} \tr (A^{\mu})^2 
\right\rangle
&=&
\frac
{\int dA \frac{1}{N} \tr (A^{\mu})^2 
\exp\left(-\frac{1}{4}\frac{1}{g_{IIB}^2} \tr ([A^M, A^N]^2) \right)}
{\int dA 
\exp\left(-\frac{1}{4}\frac{1}{g_{IIB}^2} \tr ([A^M, A^N]^2) \right)} \n
&=&
g_{IIB}
\frac
{\int dA \frac{1}{N} \tr (A^{\mu})^2 
\exp\left(-\frac{1}{4} \tr ([A^M, A^N]^2) \right)}
{\int dA 
\exp\left(-\frac{1}{4} \tr ([A^M, A^N]^2) \right)}. 
\end{eqnarray}

On the other hand, from (\ref{tune1}) and (\ref{tune2}) we obtain that
\begin{equation}
(\Lambda^{phys})^2 = g_{IIB}^{phys} 
2\pi C N^{\frac{1}{2}},
\end{equation}
which suggests that 
\begin{equation}
f(N)=2\pi C N^{\frac{1}{2}}.
\end{equation}
However, the value of $C$ is not determined in the leading order of the large N limit, because we can give any values to $\lambda$ and $\Lambda$.

Finally, we discuss how $l_s$ and $\Lambda^{phys}$ are expressed in terms of $g_{IIB}^{phys}$ and $N$. From (\ref{tune1}) we have 
\begin{equation}
l_s=C^{-\frac{1}{2}} (g_{IIB}^{phys})^{\frac{1}{2}} N^{-\frac{1}{4}}. 
\label{stringlength}
\end{equation}
Substituting (\ref{stringlength}) to (\ref{tune2}), we obtain
\begin{equation}
\Lambda^{phys}=\sqrt{2\pi}C^{\frac{1}{2}}(g_{IIB}^{phys})^{\frac{1}{2}}  N^{\frac{1}{4}}. \label{volume}
\end{equation}
These results (\ref{stringlength}) and (\ref{volume}) are expected to hold for any vacuum because the way of taking the large N limit should not depend on the vacuum. In fact, they are consistent with the results of some other analyses \cite{AIKKT}.

\vspace{1cm}

\section{Conclusion and Discussion}
\setcounter{equation}{0}

In the IIB matrix model, simultaneously diagonalizable configurations are stable and form a moduli space in the leading order of the large N limit.
If we consider fluctuations around each of them with the diagonal elements being quenched, 
we obtain a perturbative vacuum of string theory. 

Actually, type IIA string theory and type IIB string theory on $AdS_5 \times S^5$ emerge, if we consider fluctuations around appropriate configurations. 
Therefore, the moduli space should represent at least a part of perturbatively stable 
vacua of string theory.  

We have given a detailed analysis on the case of type IIA string theory.
We have shown how the string coupling $g_s$ and the string scale $l_s$ are related to the coupling constant $g_{IIB}$ and the range of the eigenvalue distribution $\Lambda$ in 
the IIB matrix model. 
As a result, we find that type IIA string theory emerges if the large N limit is 
taken with $g_{IIB}N^{-\frac{1}{2}}$ and $\Lambda N^{-\frac{1}{2}}$ being fixed. Here the freedom of overall rescaling of the matrices is fixed such that the matrices represent the string coordinates.
Any perturbative string should emerge in the same limit,  
because the way of taking the large N limit is expected not to depend on the vacuum. 
Furthermore, if we assume these relations still hold when the eigenvalue distribution 
is dynamically determined, $l_s$ and $\Lambda$ are expressed as 
\begin{eqnarray}
l_s&=&C^{-\frac{1}{2}} (g_{IIB})^{\frac{1}{2}} N^{-\frac{1}{4}} \n
\Lambda&=&\sqrt{2\pi}C^{\frac{1}{2}}(g_{IIB})^{\frac{1}{2}}  N^{\frac{1}{4}}. \label{last}
\end{eqnarray}

Let us discuss how interactions of type IIA superstring are derived in our new interpretation of IIB matrix model. The authors in \cite{2BBN} show that half-BPS classical solutions of matrix string theory determine world-sheets with definite genera when they derive the Green-Schwartz action. In this sense, the world-sheet genus expansion is not directly related to the ordinary 1/N expansion. Although correlation functions of Wilson loops factorize in the large N limit, strings are not simply represented by Wilson loops in our case. Therefore there is a possibility that interactions of strings can be reproduced only by planer diagrams. In order to examine this possibility, we need a more precise analysis on the string states, which we intend to report in future publications.

Space-times emerge in various manners in the IIB matrix model.
First, in the original picture, matrices appear as a regularization of the Schild action of 
type IIB string theory \cite{IKKT}, and they represent the space-time coordinates. 
Second, in the interpretation we have introduced in this paper, two matrices correspond to 
conjugate momenta of the world-sheet coordinates, whereas the other eight
correspond to the light-cone coordinates of type IIA string.
Third, as shortly discussed in this paper, the IIB matrix model produces the 
four-dimensional large N $\mathcal{N}=4$ super Yang-Mills theory, and thus produces 
type IIB string theory on the $AdS_5 \times S^5$ background through the AdS/CFT correspondence \cite{Maldacena}. In this case, six matrices correspond to the radial coordinate of the $AdS_5$ and the coordinates of the $S^5$, whereas the other four matrices
correspond to the conjugate momenta of the angular coordinates of the $AdS_5$. Fourth, the matrices can be regarded as the covariant derivatives on curved space-times \cite{HKK}. It is interesting to study the relations among these ways of representing space-times.

\vspace*{1cm}

\section*{Acknowledgements}
This work is supported in part by a Grant-in-Aid for the 21st Century COE ``Center for Diversity and Universality in Physics" from the Ministry of Education, Culture, Sports, Science, and Technology (MEXT) of Japan.

\end{document}